# Labour Absorption In Manufacturing Industry In Indonesia: Anomalous And Regressive Phenomena

*Tongam Sihol Nababan\*, Elvis Fresly Purba*
*Faculty of Economics and Business,  HKBP Nommensen University,*
*Medan, Indonesia*

\*Corresponding author
Email: sihol.nababan@uhn.ac.id

**Abstract**

The manufacturing industry sector was expected to generate new employment opportunities and take on labour. Gradually, however, it emerged as a menace to the sustenance of its workers. According to the findings of this study, 24 manufacturing subsectors with ISIC 2 digits in Indonesia exhibited regressive and abnormal patterns in the period 2012-2020. This suggests that, to a great extent, labour absorption has been limited and, in some cases, even shown a decline. Anomalous occurrences were observed in three subsectors: ISIC 12 (tobacco products), ISIC 26 (computer, electronic and optical products), and ISIC 31 (furniture). In contrast, regressive phenomena were present in the remaining 21 ISIC subsectors.  Furthermore, the manufacturing industry displayed a negative correlation between employment and efficiency index, demonstrating this anomalous and regressive phenomenon. This implies that as the efficiency index of the manufacturing industry increases, the index of labour absorption decreases.

JEL Classfications: J2, L6, O3, D24

## 1.  Introduction

The fundamental issue facing Indonesian employment is the incompatibility between the needs of the workforce and labour providers. This mismatch highlights a challenge in linking and matching the competencies of the workforce with industry demands. High school and university graduates possess abilities and skills that may not align with current industry requirements. The connectivity between the relevance of acquired skills and demand in the job market needs to be better established. It is evidenced that the availability of labour with digital skills does not meet the demand. Notably, there exists a significant scarcity of workers in fintech particularly in data analysis, back-end programming, user interface design, and risk management (Ferzi, 2022). This situation imposes a burden on labour not only in terms of demand, such as wage policies but also on the supply side with challenges to enhance the skills and productivity of the workforce, viewed as not aligned with the needs of the industry (Theodora, 2023).

Indonesia boasts a range of extensive and medium-scale manufacturing sectors, making it one of the country's primary industries. The manufacturing industry in Indonesia has seen remarkable growth over time, contributing significantly to national economic development, employment opportunities (Elfita and Mildawati, 2023), and the country's progress. With a population of more than 270 million, Indonesia provides ample opportunities for manufacturers to access the domestic market. Until recently, the manufacturing sector in Indonesia was predominantly characterised by labour utilisation and focused on labour-intensive production systems. However, advancements in technology and automation have led to a significant transformation in the manufacturing process, resulting in greater capital intensity and automation whilst reducing the need for labour.

The manufacturing industry is anticipated to maintain the domestic workforce through increased investment and company expansions. Labour absorption within the manufacturing sector or industry subsector entails the effective utilisation of labour resources. The comprehension of labour absorption is vital in the evaluation of the soundness and efficacy of the manufacturing industry and for informed economic and policy decision-making. However, when investment and the number of manufacturing industries increase, employment indexs decrease. Theodora (2023) noted that in 2013, an investment worth IDR 1 trillion was able to employ up to 4,594 workers. However, over time, this number has steadily decreased. In 2016, IDR 1 trillion of investment could only provide employment for 2,271 individuals. Moreover, in 2021, an investment of IDR 1 trillion can only employ 1,340 people. This situation arises from two factors: firstly, the majority of incoming investment is capital and technology-intensive, and secondly, the decline of employment in the formal sector. The surplus labour force is primarily absorbed in the informal sector.

The examination and analysis of the manufacturing industry's development is vital, providing a foundation for policymaking. A component under scrutiny is the sector's contribution to value-added output and employment, with labour functioning as an input factor for value-added production. The efficiency of labour input use and the comparison between output growth and labour absorption growth reveal the relationship between value-added contribution and labour allocation in the manufacturing industry. This study analyses the phenomenon of labour absorption in the large and medium manufacturing sector in Indonesia. Observations are conducted on the 2-digit Indonesian Standard Industrial Classification (ISIC) comprising 24 sub-sectors falling under principal classes ISIC 10 to ISIC 33.

The following text presents the labour absorption overview in the manufacturing industry in Section 2. Sections 3 and 4 compare the labour efficiency and labour absorption. Section 5 analyses the phenomenon of the relationship between output growth (value added) and labour growth. Section 6 explains the data and methods used in this study, Section 7 provides the results and discussion, and Section 8 concludes the paper.

## 2. Labour Absorption in the Manufacturing Industry

Labour absorption in the manufacturing sector refers to the employment of labour to fulfil production and operational requirements (Aisyah & Sulastri, 2020). Additionally, it can be interpreted as the number of job positions filled relative to the total population Ardiansyah et al.(2018). Labour absorption in the manufacturing sector is a crucial element of economic development and industrial growth (Pramusinto & Daerobi, 2020). These definitions describe the extent to which the manufacturing sector offers job opportunities and utilizes labour from the available workforce. Thus, high indexs of labour absorption in the manufacturing sector signify a thriving and expanding industrial sector.

Several factors are associated with labour absorption in the manufacturing industry, as previously indicated. Studies by Mundle (1983) and Seth & Seth (1991) in India have revealed a significant gap between the growth rate of output and that of labour absorption, resulting in a low net labour absorption index in the manufacturing sector. In his research, Patabendige (2004) revealed a modest decline in manufacturing sector employment, attributed to factor market distortions, labour productivity behaviour, an emerging trend towards capital intensity, real wage behaviour, and a lack of significant backward linkages. Islam (2017) discovered that in three South Asian countries (Bangladesh, India, Nepal), it is challenging to make the most of the available surplus labour without the presence of industrialisation. Consequently, the surplus labour is mainly absorbed by the service sector. In South Africa, Gumata et al. (2020) discovered that favourable employment growth shocks across all sub-sectors of manufacturing led to an increase in manufacturing employment growth, participation rates, and labour absorption, although the rates varied. The growth of employment in the manufacturing industry is more sensitive to job growth in the food, non-metallic products, basic metal products, textile, wood, and transport equipment sectors.

The research outcomes in Indonesia highlight the factors that impact labour absorption in the manufacturing sector. The study of Pramusinto & Daerobi (2020) found that variables such as wages, number of industrial enterprises, Gross Regional Domestic Product (GRDP), index of investment, technology and education have a substantial impact on labour absorption. Simanjuntak et al. (2023) conducted research in Samarinda City and Balikpapan City, which demonstrates that GRDP variables in the trade, manufacturing, and service sectors significantly impact labour absorption. However, the processing industry sector exhibits no significant influence on labour absorption. In Sulawesi Island, Alisyahbana & Anwar (2022) discovered that provincial minimum wage, number of business units, investment and GRDP all have a positive and significant influence on labour absorption in the manufacturing industry. The research done by Utami (2020) in East Java province showed that investment in medium and large manufacturing and GRDP had a favourable impact on labour absorption. Meanwhile, Amalia & Woyanti's (2020) study on medium and large industries in six provinces of Java Island displayed that investing in such industries creates a positive and significant influence on employment, whereas production value and regional minimum wage only have a positive but insignificant effect on employment.

According to the aforementioned findings, it is apparent that the manufacturing industry has a negative impact on employment rates. As noted by Manjappa (2008), Mokyr et al. (2015) (2015), and Khan & Thimmaiah (2015), capital-intensive machinery and technology used in this sector have led to low labour absorption rates.

## 3. Labour Efficiency vs Labour Absorption

Labour efficiency pertains to the effectiveness of workers in accomplishing tasks and attaining objectives within a specific timeframe Tukhtabaev (2013). Essentially, labour efficiency is focused on labour productivity, which measures the amount of output produced per unit of labour input (Bogoviz et al., 2018 ; Abukhalaf & Abusal, 2021). Efficiency is a crucial aspect of the production process that serves as a parameter in evaluating company performance (Nababan, 2019). Labour efficiency measures the extent to which a company utilises its workforce to achieve its goals. The purpose of labour efficiency is to optimise output while minimising resource consumption.

On the other side, labour providers and policymakers anticipate that manufacturing companies can hire as much labour as possible to reduce unemployment within the community. However, this can lead to a dispute as manufacturing industries strive for high productivity while still meeting employment demands from labour providers and the government. Certain factors have the potential to curtail labour use and improve production efficiency within the industry, including utilising automation techniques, implementing just-in-time manufacturing practices, establishing global supply chains, and embracing lean manufacturing strategies (Acemoglu & Restrepo, 2019 ; Javadian Kootanaee et al., 2013 ; Shih, 2020 ; Tortorella et al., 2020).

## 4. The phenomenon of the relationship between output growth (value added) and labour growth

In manufacturing, there is an anticipated correlation between output growth (value added) and labour growth, indicating that labour absorption improvements will boost output or value added. According to Hanaysha (2016), elevated productivity indexes yield superior performance and product quality. Nonetheless, this is not necessarily accurate, as it may reveal the opposite phenomenon.

Tandon (2023) asserted that the growth of the manufacturing industry must correspond to sustainable development, particularly in job creation. Nonetheless, the characteristics of the domestically available factors of production have been disregarded, leading to suboptimal outcomes in the structural transformation process and inadequate impact on employment. Dai et al.(2022) argued that the modernisation and enhancement of the labour-intensive manufacturing industry could impact the job market. A study conducted in China indicated industrial transformation and upgrading decreased job numbers, while simultaneously increasing average labour wages. Meanwhile, Perraudin et al's. (2013) research in France demonstrated that outsourcing led to the substitution of internal labour and the consequent decrease in internal employment. Meanwhile, Stryzhak (2023) suggests that the digitalization of the economy and society under the Industry 4.0 context exerts an influence on the labour market. Novel job positions with unique personnel prerequisites are being established. Meanwhile, certain fields experience fewer workers due to digitalization processes, whilst others register their count increasing. As the production digitization intensifies, the labour market's transformation process will also intensify.

Feryanto (2014) identifies noteworthy phenomena pertaining to the correlation between output growth rate and labour absorption growth rate within economic sectors in Indonesia. The first of these phenomena is categorised as an anomaly, wherein growth is high but employment growth is negative. The second is referred to as progressive, whereby output growth is lower than employment growth. The final classification is regressive, whereby output growth is higher than labour absorption growth. Fourthly, the concept of proportionality refers to when output growth is relatively balanced compared to labour absorption growth. Therefore, this study aims to investigate how this phenomenon occurs in the relationship between output growth (value added) and labour growth within the manufacturing industry in Indonesia.

## 5. Data and Methods

The study's dataset comprises the value added and the number of workers in Indonesia's large and medium manufacturing industries between 2012 and 2020. The manufacturing industry's coverage is based on the 2-digit ISIC, which includes 24 sub-sectors. Table 1 below shows the information.

Table 1
Division of Large and Medium Manufacturing Based on ISIC 2 Digits

| ISIC Code | Division of Manufacturing | ISIC Code | Division of Manufacturing |
|---|---|---|---|
| 10 | Food products | 22 | Rubber and plastic products |
| 11 | Beverages | 23 | Other non-metallic mineral products |
| 12 | Tobacco products | 24 | Basic metals |
| 13 | Textiles | 25 | Fabricated metal products except machinery and equipment |
| 14 | Wearing apparels | 26 | Computers, electronic and optical products |
| 15 | Leather and related products and footwear | 27 | Electrical equipment |
| 16 | Wood and products of wood and cork, except furniture; manufacture of articles of straw and plaiting materials, bamboo, rattan, | 28 | Machinery and equipment n.e.c |

|    | and the like |    |    |
|----|-------------|----|----|
| 17 | Paper and paper products | 29 | Motor vehicles, trailers, and semi-trailers |
| 18 | Printing and reproduction of recorded media | 30 | Other transport equipment |
| 19 | Coke and refined petroleum products | 31 | Furniture |
| 20 | Chemicals and chemical products | 32 | Other manufacturing |
| 21 | Pharmaceuticals, medicinal chemicals, and botanical products | 33 | Repair and installation of machinery and equipment |

*Source: BPS (2022), Indicator of Indonesia Manufacturing Industry 2020*

To establish the correlation between the growth rate of manufacturing value-added in ISIC subsectors (%Δ output value-added or %Δ OVA) and the rate of labour absorption (%Δ input labour absorption or %Δ ILA), the average growth rate of value-added and labour absorption can be compared. This allows for the identification of four conditions (Feriyanto, 2014):

    a. Anomaly, if %Δ OVA (+), %Δ ILA (-) ............................................... (1)
    b. Progressive, if %Δ OVA < %Δ ILA .................................................. (2)
    c. Regressive, if %Δ OVA > %Δ ILA .................................................... (3)
    d. Proportional, if %Δ OVA ∝ %Δ ILA ................................................. (4)

To analyse labour allocation efficiency in large and medium industries, we use Gasperz, (2011) proposal for the short-term Cobb-Douglas production function analysis. The necessary conditions for a short-term analysis of the Cobb-Douglas production function are: (1) total output cannot be negative (Q > 0) so the intercept coefficient (constant) must be positive (A > 0), and (2) the marginal product of the factor input must be positive, with a positive output elasticity coefficient of the input (β > 0). It is assumed that other factor inputs remain fixed, with only the amount of labour being variable.

Thus, the short-term Cobb-Douglas production function can be expressed as
$Q_{it} = AL_{it}^\beta$ ……………………………………………………………..(5)
and in logarithmic linear form, $LnQ_{it} = LnA + \beta LnL_{it} + u$ …………......(6).

Where Q represents the value-added output (in IDR) of the large and medium industry sector i in year *t*, L represents the input (number of labour) in the large and medium industry i in year *t*, and A is the constant/intercept value that demonstrates the efficiency index. The larger the value of A, the higher the efficiency of labour allocation. Parameter β represents the output elasticity of L.

Moreover, in order to examine the correlation between the expansion of labour utilization and the efficiency index derived from the Cobb-Douglas function, the OLS technique is implemented using the following model specification:

       *%Δ ILA = f (A)* ................................................................(7).

## 6. Results and Discussion

### 6.1. The phenomenon of the Relationship between the Growth Rate of Value Added and the Growth Rate of Labour Absorption

It is noteworthy that a relationship exists between the growth rate of value added in the ISIC sub-sector and the rate of labour absorption. When there is an increase in the value-added of the sub-sector, it leads to the creation of new job opportunities, resulting in labour absorption. This observation highlights an interesting phenomenon in this sector. The correlation between the growth rate of value added in ISIC sub-sectors and employment can be understood through the analysis of the mean growth rate of value-added and employment average for the years 2012-2020 outlined in Table 2 and Table 3.

Table 2.
Growth Rate of Output Value Added of Large and Medium Industries by ISIC 2 digits
Year 2012 – 2020

| ISIC Subsectors | Growth Rate of Output Value Added in Large and Medium Industries (%) | | | | | | | | |
|---|---|---|---|---|---|---|---|---|---|
| | 2013 | 2014 | 2015 | 2016 | 2017 | 2018 | 2019 | 2020 | Average |
| ISIC 10 | 33.80 | 10.30 | 8.70 | 58.90 | -5.60 | 9.70 | 4.80 | 0.20 | 15.10 |
| ISIC 11 | 52.90 | 28.50 | 31.40 | -21.00 | 61.60 | 7.20 | 12.40 | -1.20 | 21.50 |
| ISIC 12 | 47.80 | 8.50 | 25.60 | -98.00 | 3673.20 | 22.30 | 19.10 | -1.90 | 462.10 |
| ISIC 13 | 66.70 | 6.70 | 2.30 | -1.20 | 38.30 | 9.70 | 9.50 | -13.70 | 14.80 |
| ISIC 14 | 24.80 | -10.10 | 23.10 | 88.40 | 38.00 | 8.90 | 2.40 | -34.10 | 17.70 |
| ISIC 15 | 8.20 | 20.90 | 76.10 | 10.00 | 32.10 | 11.40 | 6.50 | -20.60 | 18.10 |
| ISIC 16 | 11.10 | -0.30 | 83.10 | 11.30 | 34.20 | -12.10 | -15.00 | -18.20 | 11.70 |
| ISIC 17 | 6.10 | -0.30 | -7.50 | 3.50 | 65.60 | 64.60 | -12.30 | -8.10 | 14.00 |
| ISIC 18 | 32.80 | 30.40 | 12.40 | 595.20 | -62.70 | 5.90 | -29.70 | 12.20 | 74.50 |
| ISIC 19 | 110.60 | -26.00 | 51.30 | 1316.50 | -10.10 | 42.80 | 54.40 | 10.90 | 193.80 |
| ISIC 20 | 45.40 | 15.10 | 5.20 | -14.40 | 49.60 | -6.10 | -3.20 | 10.20 | 12.70 |
| ISIC 21 | -13.70 | 29.40 | 5.10 | 110.70 | 175.10 | -22.90 | -9.80 | 1.00 | 34.40 |
| ISIC 22 | 51.10 | 55.60 | -0.40 | 12.00 | -7.30 | 9.30 | -4.40 | -21.20 | 11.80 |
| ISIC 23 | 2.80 | 69.90 | 36.60 | -17.10 | 117.60 | -29.10 | -8.90 | -15.40 | 19.60 |
| ISIC 24 | 63.00 | 4.50 | 22.50 | 7.90 | 34.90 | 28.90 | 32.60 | -0.90 | 24.20 |
| ISIC 25 | -3.50 | 2.00 | -14.40 | 88.10 | -18.50 | 14.20 | 18.70 | -7.80 | 9.90 |
| ISIC 26 | 43.70 | -3.80 | 64.30 | 23.50 | -27.50 | 0.10 | 10.50 | -18.70 | 11.50 |
| ISIC 27 | 54.60 | -6.50 | 0.60 | 61.90 | 117.20 | 3.20 | -11.90 | -31.40 | 23.40 |
| ISIC 28 | 20.60 | 42.20 | 46.80 | -41.60 | 206.80 | 8.60 | -25.30 | -15.70 | 30.30 |
| ISIC 29 | 3.30 | 14.10 | 36.00 | -41.20 | 112.30 | 0.30 | -1.50 | -18.80 | 13.10 |
| ISIC 30 | -8.70 | 13.40 | -13.20 | 136.40 | -50.70 | 37.90 | 17.20 | -35.10 | 12.20 |
| ISIC 31 | 37.00 | 96.10 | -0.90 | -31.20 | 83.20 | 10.60 | 2.30 | -12.10 | 23.10 |
| ISIC 32 | 25.20 | 37.90 | 42.70 | 31.60 | -22.20 | 24.60 | -5.30 | 27.50 | 20.20 |
| ISIC 33 | 10.40 | 19.10 | 63.90 | 75.70 | 53.40 | -22.50 | 12.10 | 5.90 | 27.30 |

Source: Results of data processing

Table 3
Growth Rate of Input Labour Absorption of Large and Medium Industries by ISIC 2 digits
Year 2012 - 2020

| ISIC Subsectors | Growth Rate of Input Labour Absorption of Large and Medium Industries (%) | | | | | | | | |
|---|---|---|---|---|---|---|---|---|---|
| | 2013 | 2014 | 2015 | 2016 | 2017 | 2018 | 2019 | 2020 | Average |
| ISIC 10 | 1.90 | -2.60 | -2.20 | 30.50 | -6.90 | -3.50 | 0.90 | -2.50 | 1.90 |
| ISIC 11 | 10.60 | 2.00 | 13.80 | 62.50 | -3.40 | -0.70 | 1.10 | -11.70 | 9.30 |
| ISIC 12 | 11.80 | -1.90 | -2.80 | -13.50 | 8.00 | -10.10 | 2.40 | -3.30 | -1.20 |
| ISIC 13 | -0.90 | 14.40 | -6.10 | 5.20 | 20.30 | -3.20 | -7.60 | -10.70 | 1.40 |
| ISIC 14 | -4.80 | 11.40 | 7.40 | 34.10 | -6.60 | -10.90 | 4.50 | -12.80 | 2.80 |
| ISIC 15 | 4.10 | 4.60 | 12.50 | 27.30 | 6.40 | -8.00 | 23.60 | -8.10 | 7.80 |
| ISIC 16 | 1.90 | -0.70 | 6.50 | 26.10 | -6.60 | -10.00 | 0.10 | -5.60 | 1.50 |
| ISIC 17 | 5.20 | 32.80 | -26.30 | 21.60 | 5.10 | -14.50 | 3.50 | -6.90 | 2.60 |
| ISIC 18 | -1.60 | -1.60 | 8.00 | 61.70 | -5.10 | -23.70 | 25.10 | -12.00 | 6.40 |
| ISIC 19 | -1.60 | -1.80 | 14.70 | 173.90 | 19.30 | 30.80 | -40.20 | 27.80 | 27.80 |
| ISIC 20 | 9.90 | -5.00 | 0.20 | 21.00 | -0.30 | -5.10 | 5.20 | 4.00 | 3.70 |
| ISIC 21 | -3.70 | -5.20 | 0.60 | 55.20 | -2.60 | 6.70 | -8.70 | 0.40 | 5.30 |
| ISIC 22 | 3.50 | 6.70 | 13.50 | 3.60 | 8.90 | -11.40 | 2.60 | -3.10 | 3.00 |
| ISIC 23 | -5.50 | -2.90 | 5.30 | 11.10 | 8.20 | -6.50 | 1.10 | -9.40 | 0.20 |
| ISIC 24 | 21.20 | 0.00 | -6.00 | 110.10 | -14.00 | 12.50 | -4.20 | 16.80 | 17.00 |
| ISIC 25 | 7.00 | -7.10 | -2.90 | 8.30 | 18.30 | -13.50 | -1.10 | 6.00 | 1.90 |
| ISIC 26 | -5.10 | -3.80 | 6.50 | -2.10 | 24.00 | -21.00 | -4.00 | -8.10 | -1.70 |
| ISIC 27 | 3.00 | 5.20 | -16.80 | 41.70 | 1.10 | 4.00 | 2.30 | -15.30 | 3.10 |
| ISIC 28 | 3.10 | 5.20 | 14.40 | 9.40 | 37.60 | -11.60 | -8.20 | 1.00 | 6.40 |
| ISIC 29 | 16.50 | 1.40 | 5.30 | 37.90 | 15.20 | 6.00 | -0.40 | 3.80 | 10.70 |
| ISIC 30 | 1.20 | 4.20 | 14.50 | 34.00 | 3.30 | -25.50 | 25.20 | -21.70 | 4.40 |
| ISIC 31 | -13.10 | 3.90 | -2.50 | 9.10 | 15.90 | -12.90 | -0.80 | -12.30 | -1.60 |
| ISIC 32 | -4.00 | 4.10 | 3.90 | 15.60 | 13.30 | -4.20 | -6.50 | 5.50 | 3.50 |
| ISIC 33 | 0.40 | 7.80 | -29.10 | 218.00 | -8.50 | -35.60 | 11.70 | -6.10 | 19.80 |

Source: Results of data processing

The growth rate of outpot value added (OVA) and the growth rate of input labour absorption (ILA) in large and medium industries in Table 2 and Table 3 are presented in Figure 1. During the period of 2012-2020, according to Figure 1, the ISIC subsector experienced anomalous and regressive phenomena, signifying low and even negative labour absorption. Anomalous phenomena were only observed in three subsectors, specifically ISIC 12 (tobacco products) with an OVA value of 4.621 and an ILA value of -0.012, ISIC 26 (computers, electronic and optical products) with an OVA value of 0.115 and an ILA value of -0.017, and ISIC 31 (furniture) with an OVA value of 0.231 and an ILA value of -0.016. The 21 remaining subsectors of ISIC display a regressive trend, where the OVA value is greater than the ILA value. As depicted in Figure 1, the ISIC subsectors with the most noteworthy employment growth rates encompass ISIC 19 (coke and refined petroleum products), ISIC 33 (repair and installation of machinery and equipment), ISIC 24 (basic metals), and ISIC 29 (motor vehicles, trailers, and semi-trailers).

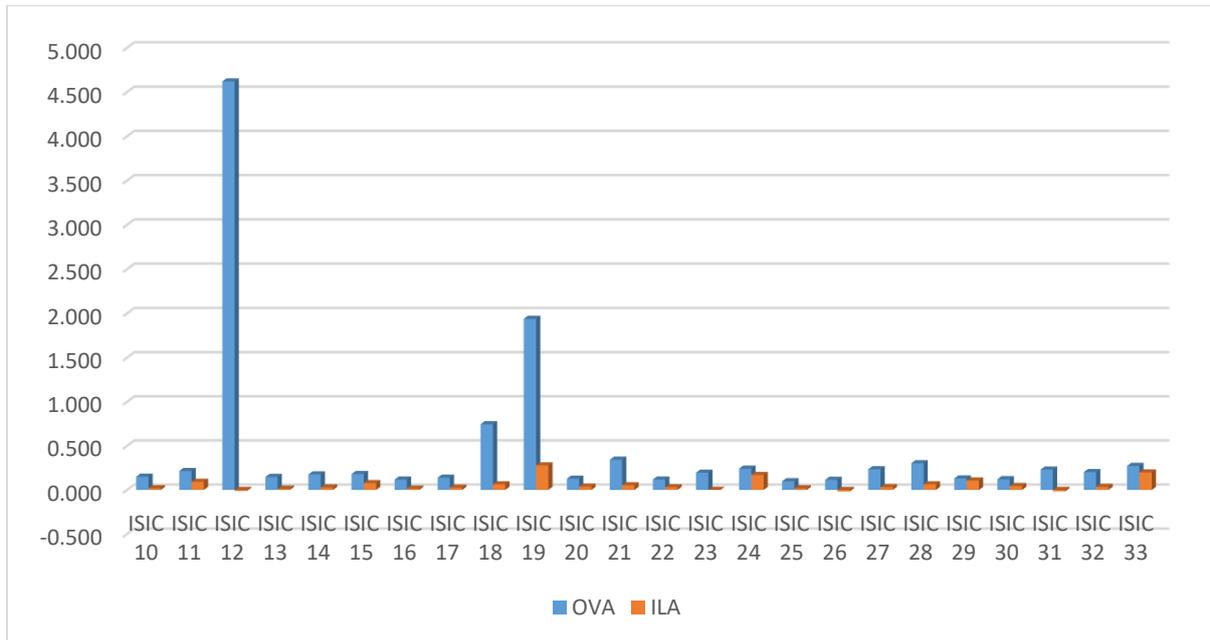

Figure 1. The growth rate of outpot value added (OVA) and the growth rate of input labour absorption (ILA) in large and medium industries

According to LIPI (2018), the rise in productivity and industrial competitiveness due to automation and robotisation poses a potential threat of job losses to human labour, as machines and robots take over. It is ironic that the manufacturing sector, which was initially expected to generate employment and provide work opportunities for labour, has now become a threat to the survival of its workforce. Consequently, such a development has led to the emergence of anomalous phenomena. However, output indexes tend to decrease due to many labourers working in low-productivity sectors. Permata et al. (2010) identified an anomalous phenomenon, as the world economy's slowdown was followed by a sharp decline in demand, resulting in a sizeable output decrease and labour rationalisation. When employees lose their jobs, they have the option to seek alternative employment within the same sector, pivot to a different sector, or pursue opportunities in the informal economy. Consequently, the rate of job placement declines significantly, potentially leading to negative growth in employment rates.

Low labour absorption may result from skill mismatches, where a notable discrepancy exists between industry-required skills and the available workforce's acquired skills. This can cause difficulty in securing qualified employees or underutilisation of skilled workers (Van der Velden & Verhaest, 2017 ; Santoso, 2015). Technology disruptions can cause low employment rates due to the adoption of new technology or automation in an industry, resulting in anomalies that displace numerous workers within a short timeframe. This raises concerns regarding labour transition and retraining (Galanakis et al.,2021). Additionally, emerging industries can also reduce employment by disrupting existing employment patterns, as noted by Okwu et al.(2022). For instance, the speedy expansion of the renewable energy industry could result in unforeseen requirements for expert competencies.

## 6.2 Relationship between Labour Absorption Rate and Efficiency Index

The linear regression equation and efficiency index are based on the Cobb-Douglass production function obtained from Section 5 formulas (5) and (6) for each ISIC subsector. The results are presented in Table 4, which shows the efficiency index of the Cobb-Douglas Production Function of KBLI Subsectors for the years 2012-2020.

Table 4.
Efficiency Index of Cobb-Douglass Production Function of ISIC 2 digits, Year 2012 - 2020

| ISIC Subsectors | Constanta | Ln L | Efficiency Index (A) | ISIC Subsectors | Constanta | Ln L | Efficiency Index (A) |
|---|---|---|---|---|---|---|---|
| ISIC 10 | -33.382 | 3.363*** | 3.180E-15 | ISIC 22 | -18.338 | 2.317*** | 1.086E-08 |
| ISIC 11 | -3.544 | 1.222*** | 2.890E-02 | ISIC 23 | -40.861 | 4.289** | 1.796E-18 |
| ISIC 12 | -21.823 | 2.607 | 3.330E-10 | ISIC 24 | -3.075 | 1.261*** | 4.619E-02 |
| ISIC 13 | -18.2146 | 2.246** | 1.229E-08 | ISIC 25 | -11.198 | 1.829 | 1.370E-05 |
| ISIC 14 | -27.041 | 2.853*** | 1.804E-12 | ISIC 26 | 11.487 | -0.055 | 9.744E+04 |
| ISIC 15 | -16.313 | 2.135*** | 8.229E-08 | ISIC 27 | -32.331 | 3.731*** | 9.096E-15 |
| ISIC 16 | -28.356 | 3.121*** | 4.843E-13 | ISIC 28 | -16.245 | 2.399*** | 8.808E-08 |
| ISIC 17 | 9.308 | 0.166 | 1.103E+04 | ISIC 29 | 3.200 | 0.731** | 2.453E+01 |
| ISIC 18 | -25.756 | 3.223*** | 6.521E-12 | ISIC 30 | -0.399 | 0.987* | 6.710E-01 |
| ISIC 19 | -14.37 | 2.541*** | 5.744E-07 | ISIC 31 | 6.207 | 3.303 | 4.962E+02 |
| ISIC 20 | -10.453 | 1.852** | 2.886E-05 | ISIC 32 | -23.783 | 2.781*** | 4.690E-11 |
| ISIC 21 | -30.477 | 3.637*** | 5.808E-14 | ISIC 33 | -5.151 | 1.381* | 5.794E-03 |
| Dependent variable : Ln Q *** ) sign. α=0.01  **) sign. α=0.05   *) sign. α=0.10 | | | | | | | |

Source: Results of data processing

For ISIC 10, the regression equation is *Ln Q = -33.382 + 3.363 Ln L*. Similarly, for ISIC 11, the regression equation is *Ln Q = -3.544 + 1.222 Ln L*, and so on. These equations are then transformed into the original Cobb-Douglas function model:

$$ISIC\ 10 : Q = e^{-33.382} L^{1.222} = (2.71828)^{-33.382} L^{3.363}$$
$$Q = (3.180E\text{-}15)L^{3.363}$$
$$ISIC\ 11 : Q = e^{-3.544} L^{0.824} = (2.71828)^{-3.544} L^{1.222}$$
$$Q = (2.890E\text{-}02) L^{1.222}$$

In the Cobb-Douglass function of ISIC 10 and ISIC 11, the coefficients A = 3.180E-15 and A = 2.890E-02 respectively indicate the efficiency index of labour use. Table 4 presents the efficiency index calculations for the remaining ISIC sub-sectors. The estimation of the Cobb-Douglass function assumes positive coefficients for both the efficiency index and elasticity index (Gasperz, 2011). Based on the table, the ISIC sub-sectors with the highest efficiency indexes are ISIC 26 (computer, electronic, and optical products), ISIC 17 (paper and paper products), ISIC 31 (furniture), and ISIC 29 (motor vehicle, trailer, and semi-trailer).

To examine the correlation between the increase in employment rate (%Δ ILA) and efficiency index, we conducted a regression analysis using the average growth rate of labour force (found in Table 3) and the efficiency index (A) (found in Table 4). The results indicate that:

*ILA = 0.056 - 0.017A\**, where the asterisk denotes significance at a 10% level.

The aforementioned regression outcome demonstrates a negative correlation between labour absorption and efficiency index. This indicates that as the efficiency index of the manufacturing industry increases, the index of labour absorption decreases. The regression result also proves that the relationship between labour absorption and efficiency level in manufacturing industry shows an anomaly and regressive phenomenon.

Certain factors can impact the efficacy of production in industry, such as automation methods, just-in-time manufacturing, global supply chains, and lean manufacturing, which aim to decrease the need for labour. Acemoglu & Restrepo (2019) posit that automation can replace tasks previously undertaken by labour, resulting in a reduction of labour. While it can bolster productivity, it may also curtail the demand for labour. The impact of automation is balanced by the emergence of new tasks where labour must have a comparative edge. According to Aghion et al. (2022), machinery and robotics are prevalent in manufacturing, which can cut down on manual labour and boost efficiency. These technologies can carry out repetitive tasks with speed and precision, resulting in increased productivity. Javadian Kootanaee et al. (2013) proposes that the implementation of just-in-time manufacturing practices can minimise inventory and reduce the need for excessive labour in managing and storing both raw materials and finished goods. Additionally, industrial efficiency is affected by global supply chains. Shih, (2020) asserts that manufacturing frequently depends on global supply chains that facilitate the procurement of materials and components from regions offering labour and production cost advantages. Lean manufacturing has been extensively implemented to enhance production efficiency. According to Tortorella et al. (2020), applying lean manufacturing principles can minimise wastage and streamline production processes, leading to more efficient utilisation of labour and resources.

## 7. Conclusions and Recommendations

Understanding labour absorption is crucial for evaluating the manufacturing industry's performance and formulating economic strategies and policies. Hence, policymakers must consider measures to sustain labour absorption as the manufacturing industry grows. A significant level of labour absorption in manufacturing signifies a vibrant and expanding industrial sector. Unfortunately, with the expansion of investment and the number of manufacturing industries, the labour absorption level declines. During the 2012-2020 period, the ISIC subsector in Indonesia exhibited anomalous and regressive occurrences. These occurrences suggest low or even negative labour absorption. The anomalous phenomenon was present in three subsectors, namely ISIC 12 (tobacco products), ISIC 26 (computer, electronic and optical products), and ISIC 31 (furniture). The remaining ISIC subsectors (21 subsectors) experienced a regressive phenomenon. This implies that as the efficiency level of the industry increases, there is a decrease in labour absorption. The regression analysis confirms the anomalous and regressive pattern of the relationship between labour absorption and efficiency level in the manufacturing industry. The correlation between labour absorption and efficiency level in the manufacturing industry is negative. The regression analysis confirms the anomalous and regressive pattern of the relationship between labour absorption and efficiency level in the manufacturing industry. The regression analysis confirms the anomalous and regressive pattern of the relationship between labour absorption and efficiency level in the manufacturing industry.

Anomalous and regressive phenomena may arise from a deceleration in the global economy, resulting in a sharp reduction in demand that precipitates a substantial decrease in output and eventually necessitates workforce rationalisation. The low labour absorption rate could stem from inadequate skill sets of the available workforce that do not fulfil industry requirements, leading to a significant mismatch. Technological disruptions may result in reduced labour absorption, as the adoption of new technologies or automation within an industry can displace a significant number of workers in a short time. The emergence of emerging industries can also disrupt the established patterns of employment. An illustrative instance is the growth of the energy sector, which has had a significant impact on the workforce.

Improving labour allocation efficiency necessitates optimizing labour resources usefulness, enhancing productivity, and remaining competitive in the market. Therefore, it requires actions such as: To effectively manage a workforce, it is essential to implement a well-crafted plan that matches labour resources with production needs. This can be achieved by matching the skills and abilities of the workforce with job roles, identifying tasks that can be automated using technology, and applying lean manufacturing principles to eliminate waste and optimise labour allocation. Additionally, a developing an enterprise resource planning system can help automate and optimise labour allocation by taking into account factors such as skills, availability, and demand.

**Funding**
This research did not receive any specific funding from public, commercial, or not-for-profit agencies.

**Declaration of Competing Interests**
To the best of our knowledge, there are no conflicts of interest or competing interests to disclose that are directly or indirectly related to this research.

**Acknowledgments**
This research was presented at the Research Results Seminar hosted by the Institute of Research and Community Service at HKBP Nommensen University on October 11th, 2023. We acknowledge all the organizers and the scientific committee of the seminar for having associated our paper with this prominent discussion.

**References**

Abukhalaf, A. H. I., & Abusal, D. 202). Measuring Labor Efficiency in Green Construction Projects. *Academia Letters*, *2*.

Acemoglu, D., & Restrepo, P. 2019. Automation and new tasks: How technology displaces and reinstates labor. *Journal of Economic Perspectives*, *33*(2), 3–30.

Aghion, P., Antonin, C., Bunel, S., & Jaravel, X. 2022. The effects of automation on labor demand: A survey of the recent literature. *SSRN (Social Science Research Network)*. https://papers.ssrn.com/sol3/papers.cfm?abstract_id=4026751

Aisyah, S., & Sulastri, S. 2020. Tracing the Labor Absorption Rate in the Medium and Large Industrial Sectors. *EcceS (Economics, Social, and Development Studies)*, *7*(2), 220–239.

Alisyahbana, A. N. Q. A., & Anwar, A. I. 2022. Determinant Analysis of Labor Absorption in the Manufacturing Industry Sector in Sulawesi Island (2010-2019). *International Conference on Social, Economics, Business, and Education (ICSEBE 2021)*, 217–223.

Amalia, D., & Woyanti, N. 2020. The Effect of Business Unit, Production, Private Investment, and Minimum Wage on the Labor Absorption in the Large and Medium Industry 6 Provinces in Java Island. *Media Ekonomi Dan Manajemen*, *35*(2), 206–217.

Ardiansyah, M., Zuhroh, I., & Abdullah, M. F. 2018. Analisis Penyerapan Tenaga Kerja Sektor Industri Pengolahan Tahun 2001-2015 di Pasuruan dan Sidoarjo. *Jurnal Ilmu

*Ekonomi*, *2*(2), 294–308.

Bogoviz, A. V, Lobova, S. V, & Ragulina, J. V. 2018. Perspectives of growth of labor efficiency in the conditions of the digital economy. *International Conference Project "The Future of the Global Financial System: Downfall of Harmony,"* 1208–1215.

Dai, Z., Niu, Y., Zhang, H., & Niu, X. 2022. Impact of the Transforming and Upgrading of China's Labor-Intensive Manufacturing Industry on the Labor Market. *Sustainability*, *14*(21), 13750.

Manjappa, DH. 2008. Productivity Performance of Selected Capital-Intensive and Labor-Intensive Industries in India During Reform Period: An Empirical Analysis. *ICFAI Journal of Industrial Economics*, *5*(4).

Elfita, A, R. and Mildawati, T. 202). "Does innovation efficiency affect financial performance? The role of ownership concentration,". *Investment Management and Financial Innovations*, *20*(1), 58–67.

Feryanto, N. 2014. *Human Resources Economics (Ekonomi Sumberdaya Manusia)*. UPP STIM YKPN.

Ferzi, N. 2022. Labour Anomalies, Large Unabsorbed Supply (Anomali Tenaga Kerja, Suplai Banyak Tak Terserap). *Https://Beritajambi.Co/Read/2022/03/04/14633/*.

Galanakis, C. M., Rizou, M., Aldawoud, T. M. S., Ucak, I., & Rowan, N. J. (2021). Innovations and technology disruptions in the food sector within the COVID-19 pandemic and post-lockdown era. *Trends in Food Science & Technology*, *110*, 193–200.

Gasperz, V. 201). *Managerial Economics (Ekonomi Manajerial)*. Gramedia Pustaka Utama.

Gumata, N., Ndou, E., Gumata, N., & Ndou, E. 2020. What Is the Impact of the Manufacturing Sector Output and Sub-sector Employment Growth on the Labour Absorption and Participation Rates? *The Secular Decline of the South African Manufacturing Sector: Policy Interventions, Missing Links and Gaps in Discussions*, 301–309.

Hanaysha, J. 2016. Improving employee productivity through work engagement: Evidence from higher education sector. *Management Science Letters*, *6*(1), 61–70.

Islam, R. 2017. Structural transformation and absorption of surplus labour. *The Bangladesh Development Studies*, *40*(3 & 4), 105–135.

Javadian Kootanaee, A., Babu, K. N., & Talari, H. 2013. Just-in-time manufacturing system: from introduction to implement. *Available at SSRN 2253243*.

Khan, S., & Thimmaiah, N. 2015. Economic reforms and sources of productivity growth in selected organised manufacturing labour intensive and capital intensive industries in India-a comparative study. *Economic Affairs*, *60*(2), 301.

LIPI. 2018. *Industrial Revolution 4.0: Dilemmas, Inequalities and Anomalies, Monthly Discussion (Revolusi Industri 4.0: Dilema, Ketimpangan Dan Anomali, Diskusi Bulanan)*. LIPI. https://kependudukan.brin.go.id/berita/revolusi-industri-4-0-dilema-ketimpangan-dan-anomali

Mokyr, J., Vickers, C., & Ziebarth, N. L. 2015. The history of technological anxiety and the future of economic growth: Is this time different? *Journal of Economic Perspectives*, *29*(3), 31–50.

Mundle, S. 1983. Labour Absorption in Agriculture and Restricted Market for Manufacturing

Industry: An Aspect of Long-Term Consequences of Colonial Policy in Asia. *Economic and Political Weekly*, 767–778.

Nababan, T. S. 2019. Efficiency and Elasticity of Labor Use on Economic Sectors in Indonesia. *Advances in Economics, Business and Management Research*, *86*.

Okwu, M. O., Tartibu, L. K., Maware, C., Enarevba, D. R., Afenogho, J. O., & Essien, A. 2022. Emerging Technologies of Industry 4.0: Challenges and Opportunities. *2022 International Conference on Artificial Intelligence, Big Data, Computing and Data Communication Systems (IcABCD)*, 1–13.

Patabendige, A. J. 2004. *Trends in the factor markets and their effects on labour absorption: A study on the Sri Lankan manufacturing industry*. The University of Waikato.

Permata, M. I., Yanfitri., & Prasmuko, A. 2010. Labor Shifting Phenomenon in the Indonesian Labor Market (Fenomena Labor Shifting Dalam Pasar Tenaga Kerja Indonesia). *Buletin Ekonomi Moneter Dan Perbankan*, *January 20*. https://www.bmeb-bi.org/index.php/BEMP/article/download/243/220/

Perraudin, C., Thèvenot, N., & Valentin, J. 201). Avoiding the employment relationship: Outsourcing and labour substitution among French manufacturing firms, 1984–2003. *International Labour Review*, *152*(3–4), 525–547.

Pramusinto, N. D., & Daerobi, A. 202). Labor Absorption of the Manufacturing Industry Sector in Indonesia. *Budapest International Research and Critics Institute-Journal (BIRCI-Journal)*, *3*(1), 549–561.

Santoso, G. 201). Technology as a driver of skills obsolescence and skills mismatch: Implications for the labour market, society and the economy. *ANU Undergraduate Research Journal*, *7*, 49–62.

Seth, V. K., & Seth, A. K. 191). Labour absorption in the Indian manufacturing sector. *Indian Journal of Industrial Relations*, 19–38.

Shih, W. C. 2020. Global supply chains in a post-pandemic world. *Harvard Business Review*, *98*(5), 82–89.

Simanjuntak, F., Suharto, R. B., Lestari, D., & Zulfikar, A. L. 2023. The Influence of the Gross Regional Domestic Product of the Trade and Manufacturing Industry and Services Sector on the Amount of Labor Absorption in the City of Samarinda and the City of Balikpapan. *International Journal of Management Research and Economics*, *1*(3), 69–74.

Stryzhak, O. 2023. Analysis of Labor Market Transformation in the Context of Industry 4.0. *Studia Universitatis Vasile Goldiş, Arad-Seria Ştiinţe Economice*, *33*(4), 23–44.

Tandon, A. (2023). *Labour and Capital Use in Indian Manufacturing: Structural Aspects*. Taylor & Francis.

Theodora, A. 2023. Investment Anomaly, Growing High but Not Absorbing Many Workers (Anomali Investasi, Tumbuh Tinggi tetapi Tak Banyak Menyerap Pekerja). *Https://Www.Kompas.Id/Baca/Ekonomi/2023/02/04*.

Tortorella, G., Cómbita-Niño, J., Monsalvo-Buelvas, J., Vidal-Pacheco, L., & Herrera-Fontalvo, Z. 2020. Design of a methodology to incorporate Lean Manufacturing tools in risk management, to reduce work accidents at service companies. *Procedia Computer Science*, *177*, 276–283.

Tukhtabaev, J. 2013. Criteria and parameters of labor efficiency. *Association 1901" SEPIKE*,

201.

Utami, B. S. A. 2020. Analisis Penyerapan Tenaga Kerja Pada Sektor Industri Manufaktur (Besar dan Sedang) Propinsi Jawa Timur. *Journals of Economics Development Issues (JEDI)*, *3*(1), 21–30.

Van der Velden, R., & Verhaest, D. 2017. Are Skill Deficits always Bad? Toward a Learning Perspective on Skill Mismatches☆. In *Skill mismatch in labor markets* (pp. 305–343). Emerald Publishing Limited.